\documentclass[twocolumn,showpacs,preprintnumbers,amsmath,amssymb,prb]{revtex4}
\usepackage[dvips]{color,graphics,epsfig,rotating}
\usepackage{graphicx}
\usepackage{dcolumn}
\usepackage{bm}

\begin{document}

\title{Electron Phonon Superconductivity in LaNiOP}

\author{Alaska Subedi}
\affiliation{ Department of Physics and Astronomy, University of
Tennessee, Knoxville, Tennessee 37996-1200, USA}
\affiliation{Materials Science and Technology Division,
Oak Ridge National Laboratory, Oak Ridge, Tennessee 37831-6114} 

\author{D.J. Singh}
\affiliation{Materials Science and Technology Division,
Oak Ridge National Laboratory, Oak Ridge, Tennessee 37831-6114} 

\author{M.-H. Du}
\affiliation{Materials Science and Technology Division,
Oak Ridge National Laboratory, Oak Ridge, Tennessee 37831-6114} 

\date{\today} 

\begin{abstract}
We report first principles calculations of the electronic structure,
phonon dispersions and electron phonon coupling of LaNiPO.
These calculations show that this material can be explained as
a conventional electron phonon superconductor in contrast to the
FeAs based high temperature superconductors.
\end{abstract}


\maketitle

The discovery of high temperature superconductivity in a series
of FeAs based compounds, prototype LaFeAs(O,F) by Kamihara
and co-workers \cite{kamihara}
has led to intense investigation of these phases
and the discovery of a number of new phases with critical temperatures
in some cases near and above 55K. \cite{ren,wang,kito}
In addition, recently, high temperature
superconductivity with similar properties
has been found in simpler compounds based on FeAs sheets such as
ThCr$_2$Si$_2$ structure type (Ba,K)Fe$_2$As$_2$.
\cite{rotter1,rotter2}
These Fe-based materials are commonly viewed as unconventional
(non-electron-phonon) superconductors based on the high values of $T_c$,
the proximity to magnetism and the fact that calculations of the
electron phonon coupling show that it is far too weak to account
for the superconductivity. \cite{boeri,mazin}
It should be emphasized that these are rich structure types
and that there is no doubt much compositional space that remains to be
explored. \cite{quebe,pfisterer}

In this regard,
besides the FeAs based phases, superconductivity is also observed
in Ni based materials, including LaNiPO, \cite{watanabe,tegel}
pure, fluorine or Sr doped LaNiAsO, \cite{li,fang,chen}
and LaNiBiO. \cite{kozhevnikov}
Interestingly, both Ni and Fe are ambient temperature elemental
ferromagnets, and many Fe and Ni compounds show magnetism.
In addition, it is interesting to note a possible
connection to the cuprates. Specifically, the Ni compounds
are based on square lattices of nominally Ni$^{2+}$ and superconduct
with various dopings, including electron doping
to yield a nominal $d$ occupancy 8+$\delta$, while the
high $T_c$ superconductors are based on square lattices of Cu$^{2+}$
with hole doping for nominal $d$ occupancy 9-$\delta$, where $\delta$
is the doping level.

Returning to the connection with the Fe-based compounds, both
electronic structure calculations \cite{singh-du,zhang,shein1,shein2}
and experiment show important
differences between the Ni-based and Fe-based superconductors.
In particular, the Fe-based materials show low carrier density
with relatively small Fermi surfaces,
high density of states and proximity to magnetism,
\cite{mazin,singh-du,ishibashi,yildirim}
while the Ni compounds show large Fermi surfaces, lower density of
states and are apparently further from magnetic instabilities.
This suggests that the Ni-based compounds may be a different class
of superconductors, perhaps, considering the lower $T_c$ observed
so far in this group, conventional electron-phonon materials.
On the other hand, it may be noted that the kinds of measurements
and calculations done so far would have given qualitatively
similar results for cuprate superconductors in the the optimal and
over-doped regions - i.e. moderate density of states, high carrier
density metals, apparently far from magnetism.

Here we report details of our previous electronic structure
calculations as well as calculations of the phonon dispersions
and electron-phonon
coupling for LaNiPO. The calculations show that unlike the Fe-based
materials, superconductivity in this Ni-based compound is readily explained
by the standard electron-phonon mechanism.
This means that the superconductivity of LaNiPO is not related to that
of the FeAs based materials.

\begin{figure}
\includegraphics[height=3.2in,angle=270]{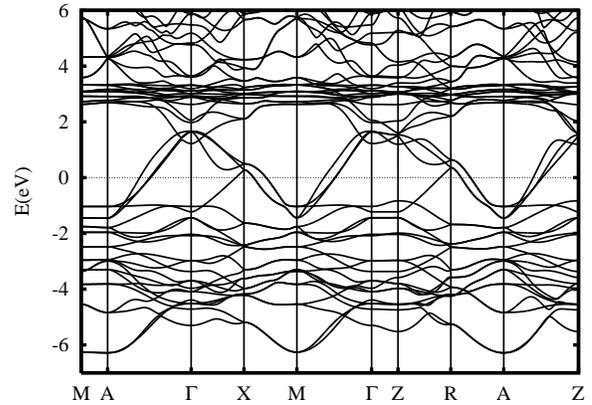}
\caption{\label{lda-bands}
Calculated LDA band structure of LaNiPO using the experimental lattice
parameter and relaxed internal coordinates,
$z_{\rm La}$=0.1506 and $z_{\rm P}$=0.6210.
}
\end{figure}

\begin{figure}
\includegraphics[height=3.2in,angle=270]{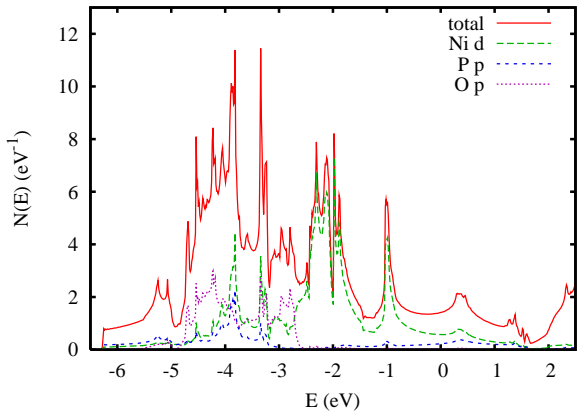}
\caption{\label{lda-dos} (color online)
Calculated LDA density of states of LaNiPO on a per formula unit both spins
basis. The projections are onto the LAPW spheres.
}
\end{figure}

\begin{figure}
\includegraphics[height=3.2in,angle=270]{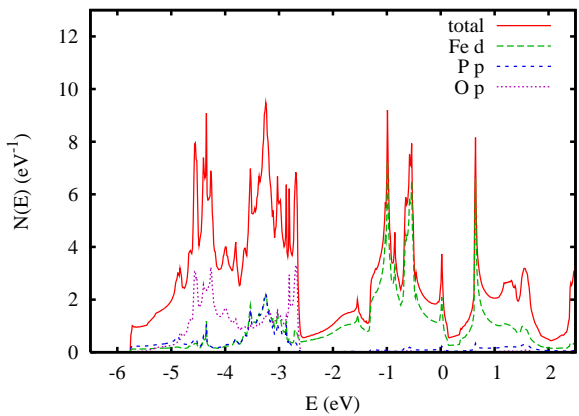}
\caption{\label{fe-dos} (color online)
Calculated LDA density of states of LaFePO on a per formula unit both spins
basis. The projections are onto the LAPW spheres. The Fe sphere radius was
2.1 $a_0$, which is equal to the Ni sphere radius used in Fig. \ref{lda-dos}.
}
\end{figure}

\begin{figure}
\includegraphics[width=3.0in,angle=0]{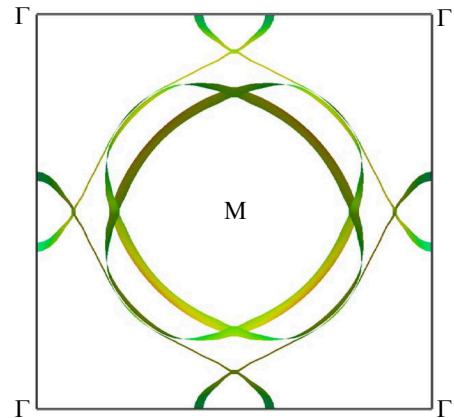}
\includegraphics[width=3.2in,angle=0]{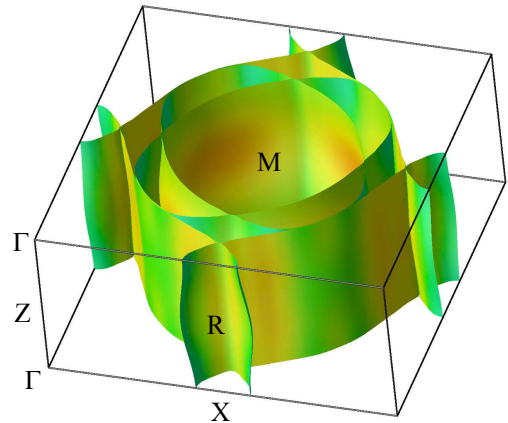}
\caption{\label{fermi} (color online)
Calculated LDA Fermi surface of LaNiPO shaded by velocity.
The top panel shows a view along the tetragonal axis while
the bottom panel shows a tilted view.
}
\end{figure}

Our electronic structure was discussed briefly previously
\cite{singh-du} and is similar to that reported by Zhang and
co-workers for the same compound. \cite{zhang}
It also shows similarities to the electronic structures found for the
other Ni-based compounds. \cite{shein1,shein2}
Our electronic structure calculations were performed within the local
density approximation (LDA) using the general potential linearized
augmented planewave (LAPW) method \cite{singh-book}
as for LaFeAsO (Ref. \onlinecite{singh-du}).
LAPW sphere radii of 2.2 $a_0$, 2.1 $a_0$, 2.1 $a_0$ and 1.6 $a_0$
were used for La, Ni, P, and O, respectively.
In particular we used the experimentally reported
tetragonal lattice parameters ($a$=4.0461\AA, $c$=8.100\AA) \cite{watanabe}
and relaxed the internal coordinates
which correspond to the La and P heights.
We obtain
$z_{\rm La}$=0.1506 and $z_{\rm P}$=0.6210,
which are close to the reported experimental values of
$z_{\rm La}$=0.1531 and $z_{\rm P}$=0.6260, \cite{watanabe}
and
$z_{\rm La}$=0.1519 and $z_{\rm P}$=0.6257. \cite{tegel}
This is different from the Fe-based compounds where pnictogen
heights significantly lower than the reported experimental
values are generally obtained. \cite{yin,mazin-johannes}

The calculated band structure and electronic density of states (DOS)
are shown in Figs. \ref{lda-bands} and
\ref{lda-dos}, respectively. The DOS for LaFePO, calculated
in the same way is shown in Fig. \ref{fe-dos} for comparison.
The Fermi surface of LaNiPO is given in Fig. \ref{fermi}.
As may be seen, these
are very different in LaNiPO from those in LaFeAsO.
This is as might be expected from the different electron count. 
LaNiPO and LaFeAsO have a similar structure to the $d$ bands,
and in particular LaNiPO should be described as
Ni$^{2+}$ ions on a square lattice with direct hopping as well
as a P induced crystal field.
In fact, the projections of the density of states shows that
there is a greater degree of covalency between Ni and P,
than between Fe and As or Fe and P in the corresponding Fe-based
materials.
There remains a pseudogap at a $d$ electron count of 6, however
Ni$^{2+}$ has 8 electrons.
This places $E_F$ well inside
the upper manifold of $d$
states.
In this energy range the bands are derived
from Ni $d$ states hybridized with P $p$ states.
The bands at $E_F$ are more dispersive than in the
Fe compounds where $E_F$ is lower.
This leads to lower density of states with higher in plane Fermi velocity,
$N(E_F)$=1.41 eV$^{-1}$ per formula unit, $v_{xx}$=3.75x10$^7$ cm/s,
$v_{zz}$=0.39x10$^7$ cm/s.
The lower $N(E_F)$ puts the Ni based
compound further from magnetism than the
Fe-based materials, as was discussed. \cite{singh-du}
Furthermore, this compound is quite two dimensional in the sense that
there is no 3D Fermi surface sheet. Based on the anisotropy of the
Fermi velocity, the ratio of in-plane to $c$-axis conductivity for the
Ni compound is $\sim$ 100 assuming isotropic scattering.
The Fermi surface may described as consisting of two large ellipsoidal
cross-section electron cylinders around M, and a large hole
section around $\Gamma$. This hole section intersects the zone
boundary near $X$ leading to an electron section around $X$.

\begin{figure}
\includegraphics[width=3.0in,angle=0]{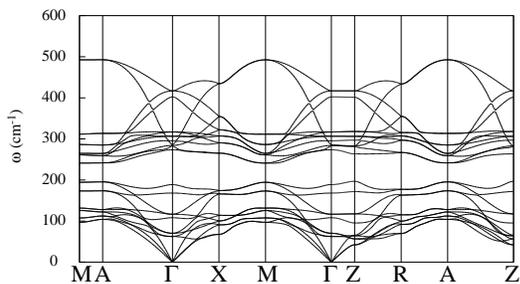}
\caption{\label{phonons}
Calculated phonon dispersions of LaNiPO as obtained within linear response.
}
\end{figure}

\begin{figure}
\includegraphics[height=3.0in,angle=270]{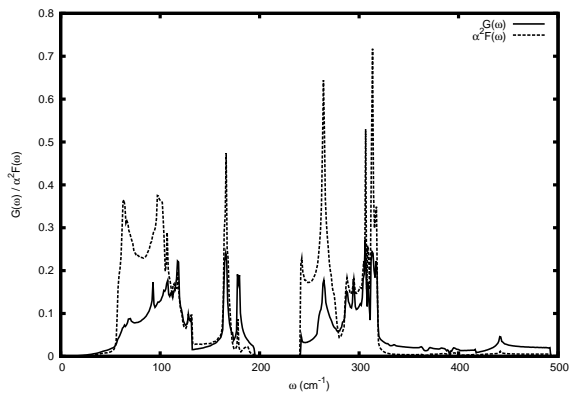}
\caption{\label{a2f}
Calculated phonon density of states and electron phonon spectral
functional $\alpha^2 F(\omega)$.
}
\end{figure}

The phonon and electron phonon calculations were performed in linear
response, with the quantum espresso code \cite{qe} and
ultrasoft pseudopotentials and the generalized gradient approximation
of Perdew, Burke and Ernzerhof (PBE), \cite{pbe}
similar to our previous calculations for LaFeAsO. \cite{mazin}
We did convergence tests for the basis set size, the planewave
expansion of the charge density, the temperature broadening,
and the Brillouin zone sampling.
An 8x8x4 grid was used for the zone integration
in the phonon calculations, while a more dense 32x32x8 grid was used
for the zone integration in the electron-phonon coupling calculation.
The basis set cut-off for the wavefunctions was 40 Ry, while a 400 Ry
cut-off was used for the charge density.

The calculated phonon dispersions of LaNiPO are shown in Fig. \ref{phonons}.
The corresponding phonon density of states and Eliashberg spectral function
$\alpha^2 F(\omega)$
are shown in Fig. \ref{a2f}.
The phonon dispersions show a set of 12 phonon bands extending up to
$\sim$ 200 cm$^{-1}$, separated by a gap from 12 higher frequency bands
extending up to $\sim$ 500 cm$^{-1}$ (note that there are 24 phonon
branches since there are two formula units per primitive cell).
The higher frequency manifold is derived mainly from O and P motions.
Within this upper manifold the P contribution is mainly below
$\sim$ 300 cm$^{-1}$, while the dispersive modes above 300 cm$^{-1}$
are mainly O derived.
The lower manifold from 0 to 200 cm$^{-1}$ consists of the acoustic
modes and modes of mixed, but mainly metal character.

We obtain a value of the electron phonon coupling $\lambda$=0.58
with logarithmically averaged frequency $\omega_{\rm ln}$=113 cm$^{-1}$.
Relative to the phonon density of states, the spectral function is
enhanced for the lower frequency metal modes, which have
strong in-plane Ni character,
and for the modes at the bottom of the upper manifold, which have
strong P character.
Thus in spite of the lower electronic density of states, we obtain
stronger electron phonon coupling as compared with LaFeAsO, where
$\lambda$ $\sim$ 0.2. \cite{boeri,mazin}
Inserting these numbers into the simplified Allen-Dynes formula,

\begin{equation}
k_B T_c = {{\hbar \omega_{\rm ln}} \over {1.2}} ~~ {\rm exp}
\left \{
- {{1.04 (1 + \lambda)} \over {\lambda - \mu^* (1 + 0.62 \lambda)}}
\right \} ,
\end{equation}

with ordinary values of the Coulomb parameter $\mu^*$ yields
values in reasonable accord with experiment. Specifically,
for $\mu^*$=0.12 we obtain $T_c$=2.6K,
which is in accord with the experimental value $T_c$ $\sim$ 3K,
\cite{watanabe}
or $T_c$=4.2K.
\cite{tegel}

In conclusion we find that LaNiPO has a conventional
superconducting state which arises from
band metal with moderate density of states and intermediate
electron phonon coupling. This is in contrast to e.g.
LaFeAs(O,F) which is a high density of states material, near magnetism,
and with weak electron phonon coupling that can in no way explain the
superconductivity.
This leaving aside structure and chemistry,
the superconductivity of LaNiPO and presumably
the rest of the Ni-based oxypnictides is unrelated
to that of the Fe-based materials.

We are grateful for discussions with I.I. Mazin,
D.G. Mandrus, B.C. Sales, R. Jin,
and M. Fornari and support from DOE,
Division of Materials Sciences and Engineering.

\end{document}